\documentclass[aps,groupedaddress]{revtex4}
\usepackage{color,graphicx}

\begin{document}
\definecolor{MyDarkGreen}{rgb}{0,0.45,0}
\definecolor{MyDarkBlue}{rgb}{0,0,0.75}
\definecolor{MyDarkRed}{rgb}{0.9,0,0}
\definecolor{MyDarkPurple}{rgb}{0.45,0,0.45}
\title{Critical power of collapsing vortices}
\author{Gadi Fibich}
\email[]{fibich@tau.ac.il} \homepage[]{www.math.tau.ac.il/~fibich}
\affiliation{School of Mathematical Sciences, Tel Aviv University,
Tel Aviv 69978, Israel}
\author{Nir Gavish}
\email[]{nirgvsh@tau.ac.il} \homepage[]{www.tau.ac.il/~nirgvsh}
\affiliation{School of Mathematical Sciences, Tel Aviv University,
Tel Aviv 69978, Israel}
\begin{abstract}We calculate the critical power for collapse of linearly-polarized phase vortices,
and show that this expression is more accurate than previous
results. Unlike the non-vortex case, deviations from radial symmetry
do not increase the critical power for collapse, but rather lead to
disintegration into collapsing non-vortex filaments. The cases of
circular, radial and azimuthal polarizations are also considered.
\end{abstract}

 \maketitle

 The nonlinear optical process of
self-focusing sets an upper limit on the amount of laser power that
can be propagated through a medium with an intensity dependent
refractive index (i.e., $n = n_0 + n_2I$, where $n_0$ is the linear
refractive index, $n_2$ is the nonlinear refractive index, and I is
the intensity). For powers above this threshold the beam will
undergo collapse, with the peak intensity becoming sufficiently high
that damage to the material can occur.  Ultimately, collapse will be
arrested by some physical mechanism, such as plasma formation,
normal dispersion or damping.

Let us briefly review the situation in the non-vortex case.  The
value of the critical power is given by~\cite{critical-00}
\[P_{cr} = \frac{\lambda^2}{4\pi n_0n_2}p_{cr} ,\]
where $p_{cr}$ is the non-dimensional critical power for collapse in
the dimensionless NLS
\begin{equation}
i\psi_z(z,x,y)+\Delta\psi+|\psi|^2\psi=0,\quad \psi(0,x,y)=\psi_0(x,y). 
\label{eq:NLS}
\end{equation}
In the NLS model, there is no mechanism for arrest of collapse,
hence collapse is defined as the maximal amplitude becoming
infinite. Weinstein~\cite{Weinstein-83} proved that the lower bound
for the critical power is equal to~$p_{cr}=\int |R|^2\,rdr\approx
1.86$, i.e., the power of the {\em Townes profile}, which is the
ground state solution of
\[
R''+\frac1r R'-R+R^3=0,\qquad R'(0)=0,\qquad R(\infty)=0.
\]
While the Townesian input beams~$\psi_0=\lambda R(\lambda r)$,
where~$\lambda>0$, can collapse with exactly the input
power~$p_{cr}$, all other input profiles require power strictly
above~$p_{cr}$ for collapse~\cite{Merle-92,Merle-93}. In practice,
however, the critical power of peak-type (i.e., non ring-type)
radially-symmetric input beams is only a few percents
above~$p_{cr}$~\cite{critical-00,Elliptic-00}. For example, the
critical power of Gaussian and super-Gaussian ($\psi_0=c\,e^{-r^4}$)
input beams is~$\approx2\%$ and~$\approx 8\%$ above~$p_{cr}$,
respectively.

We now consider the critical power of vortex input beams.
In~\cite{Kruglov-92}, Kruglov et al. derived an expression for the
critical power of vortex beams, and showed that it increases with
the winding number (or topological charge)~$m$. In this study, we
show that this expression is inaccurate, and derive the correct
expression for the critical power. Unlike the vortex-free case,
deviations from radial-symmetry do not increase the critical power,
but rather lead to disintegration into collapsing non-vortex
filaments.

\begin{figure}
\begin{center}
\scalebox{1}{\includegraphics{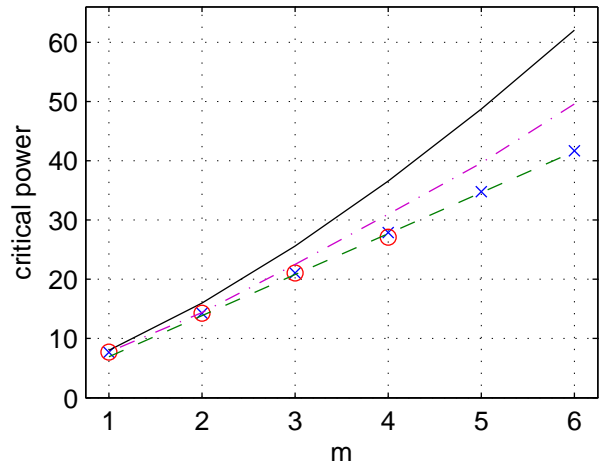}}
\caption{(color online) Critical power $p_{cr}(m)=\int_0^\infty
R_m^2\,rdr$~(\textcolor{MyDarkBlue}{$\times$}), the approximation~$4\,\sqrt3 m$
(\textcolor{MyDarkGreen}{dashed line}),
numerical estimate of the critical power as a function
of~$m$~(\textcolor{MyDarkRed}{$\circ$},
data taken from~\cite{Kruglov-92}), the critical power for collapse of Laguerre-Gaussians~(\textcolor{MyDarkPurple} {dash-dotted line}) and the analytic
 estimate~$I_{cr}^{(m)}$ (\cite{Kruglov-92}, solid line). }
\label{fig:match}
\end{center}
\end{figure}

We first consider radially-symmetric vortex input beams of the
form~$\psi_0=A_0(r)e^{im\theta}$.  In this case, the solution
remains a vortex with winding number m, i.e., it is of the
form~$\psi(z,r,\theta)=A(z,r)e^{im\theta}$~\cite{vortex-08}.
Following a similar derivation to~\cite{Weinstein-83}, it can be
rigorously shown that the lower bound for the critical power of
radially-symmetric vortex input beams~$\psi_0=A_0(r)e^{im\theta}$ is
\[p_{cr}(m)=\int |R_{m}|^2\,rdr,\] where $R_{m}$ is the ground state
solution of
\[
R_m''(r)+\frac1r
R_m'-\left(1+\frac{m^2}{r^2}\right)R_m+R_m^3=0,\quad
R_m'(0)=0,\quad R_m(\infty)=0.
\]
The values of~$p_{cr}(m)$ for~$m=1,\cdots,6$ are listed in
Table~\ref{tab:critical_Power}.  Using the
approximation~\cite{Mizumachi-07}
\begin{equation}\label{eq:Rm_app}
R_{m}(r)\approx\sqrt3\mbox{sech}\left(\frac{r-\sqrt2m}{\sqrt{2/3}}\right),
\end{equation}
we can derive the analytic approximation~$p_{cr}(m)\approx4\sqrt3m$.
Figure~\ref{fig:match} shows that~$p_{cr}(m)$ is well approximated
by~$4\sqrt3\,m$, and that the approximation improves as~$m$
increases.
\begin{table}\caption{}
\begin{center}
\begin{tabular}{|c||c|c|c|c|c|c|}\hline
m&1&2&3&4&5&6\\\hline
$p_{cr}(m)/p_{cr}$&4.12&7.65&11.3&15.0&18.7&22.4
\\\hline
\end{tabular}
\label{tab:critical_Power} \end{center}
\end{table}

We now consider the critical power of various vortex input profiles,
and ask under what condition the critical power is close to the
lower bound~$p_{cr}(m)$.  As in the vortex-free case, the only input
profiles that can collapse with input power exactly equal
to~$p_{cr}(m)$ are~$\psi_0=\lambda R_{m}(\lambda r)e^{im\theta}$. We
first calculate the critical power of the Laguerre-Gaussians
profiles
\[\psi_0^{LG}=c\,r^me^{-r^2}e^{im\theta},\]
which are the vortex modes of the linear Schr\"odinger equation. To
do that, we solve the NLS with the initial condition~$\psi_0^{LG}$
and gradually increase~$c $ until, at~$c_{th}$, the solution
collapses. In this case, the critical power is close to~$p_{cr}(m)$
for~$m=1$ but as~$m$ increases, the excess power above~$p_{cr}(m)$
needed for collapse increases, see
Table~\ref{tab:Power_approximation}. Similarly, for the sech input
profile
\[
\psi_0^{sech}=c\,r^2\mbox{sech}(r-5)e^{im\theta}, \label{eq:sech_r=5}
\]
the critical power is close to~$p_{cr}(m)$ only for~$m=2,3,4$, see
Table~\ref{tab:Power_approximation}.

To better understand these results, let us consider vortex profile
of the form~$\psi_0=c f(r)e^{im\theta}$, where
\[
f(r)=Q(\rho),\qquad
\rho=\frac{r-r_{max}}{L},
\]
and~$Q(\rho)$ attains its maximum at $\rho=0$. This ring profile is
characterized by the ring width~$L$ and radius~$r_{max}$.   As in
the vortex-free case, the closer~$f$ is to a member of the
one-parameter family~$\lambda R_{m}(\lambda r)$, the smaller the
excess power above~$p_{cr}(m)$ needed for collapse.
By~(\ref{eq:Rm_app}), the family~$\lambda R_{m}(\lambda r)$ is
characterized by
\begin{equation}
\mbox{radius/width}=\sqrt3m.\label{eq:radius_width}
\end{equation}
Therefore, $f(r)$~has to satisfy~(\ref{eq:radius_width}) to
``leading order'' to be close to~$\lambda R_{m}(\lambda r)$.

The Laguerre-Gaussian modes~$\psi_0^{LG}$ are characterized
by~$\mbox{radius}/\mbox{width}=\sqrt{m/2}$. This ratio is close
to~(\ref{eq:radius_width}) only for~$m\approx1$, explaining why the
critical power of Laguerre-Gaussian modes is close to~$p_{cr}(m)$
only for~$m=1$. Similarly, the sech profile~$\psi_0^{sech}$ is
characterized by~$\mbox{radius}/\mbox{width}=5$. Since the
radius/width of~$\lambda R_{m}(\lambda r)$ is equal to~$\sqrt{3}m$,
this ratio is close to $5$ for~$m=\frac{5}{\sqrt3}\approx 2.88$, see
equation~(\ref{eq:radius_width}).  This explains why the threshold
power of the sech profile~$\psi_0^{sech}$ is closest to~$p_{cr}(m)$
for~$m=3$.  As a final confirmation of this observation, we ``fix''
the sech profile~$\psi_0^{sech}$ so that ``it behaves like
a~$\lambda R_{m}(\lambda r)$ profile'', i.e., that it
satisfies~(\ref{eq:radius_width}) to leading order, as follows:
\begin{equation}
\psi_0^{m-sech}=
\sqrt2\left(\frac{r}{\sqrt3m}\right)^2\mbox{sech}\left(r-\sqrt
3m\right)e^{im\theta}\label{eq:corrected_sech}.
\end{equation}
Indeed, the threshold power of the ``modified'' sech
profile~(\ref{eq:corrected_sech}) is less than~$1\%$ above the
critical power  for~$m=1,\cdots,6$, see
Table~\ref{tab:Power_approximation}.

\begin{table}\begin{center}\caption{Excess power above~$p_{cr}(m)$ needed for collapse.}
\begin{tabular}{|c|c|c|c|}\hline
m&\multicolumn{3}{c|}{Input beams}\\\hline
&$\psi_0^{LG}$&$\psi_0^{sech}$&$\psi_0^{m-sech}$
\\\hline 1&0.65\%&20\%&0.13\%
\\\hline 2&0.80\%&4.5\%&0.91\%
\\\hline 3&7\%&1.9\%&0.71\%
\\\hline 4&11\%&2.9\%&0.32\%
\\\hline 5&14\%&9\%&0.17\%
\\\hline 6&19\%&14\%&0.34\%
\\\hline
\end{tabular}
\label{tab:Power_approximation} \end{center}
\end{table}
In~\cite{Kruglov-92}, Kruglov et al. estimated the critical power
for vortex collapse to be equal to
\begin{equation}
I_c^{(m)}=\frac{2^{2m+1}m!(m+1)!}{(2m)!}.
\label{eq:Icm}
\end{equation}
In~\cite{Kruglov-92}, they also estimated the critical power
numerically for $m=1,2,3$ and $4$.  These numerical results agree
with our analytic calculation of~$p_{cr}(m)$, but not with their own
estimate~$I_c^{(m)}$, see Figure~\ref{fig:match}. To understand why
this is the case, we note that the derivation of~$I_{c}^{(m)}$ was
based on the assumption that the collapsing vortex has a
self-similar Laguerre-Gaussian profile. As noted before, the
Laguerre-Gaussian modes are not a good approximation of the
one-parameter family~$\lambda R_{m}(\lambda r)$, and
as~$m$~increases this approximation becomes less and less accurate.
In addition, the assumption that the solution undergoes an
aberrationless (adiabatic) self-similar collapse is known to lead to
over-estimates of the critical power~\cite{Elliptic-00}. Indeed,
even for Laguerre-Gaussian input beams, the critical power is closer
to~$p_{cr}(m)$ than to~$I_{cm}(m)$, see Figure~\ref{fig:match}.

\begin{figure}
\begin{center}
\scalebox{1}{\includegraphics{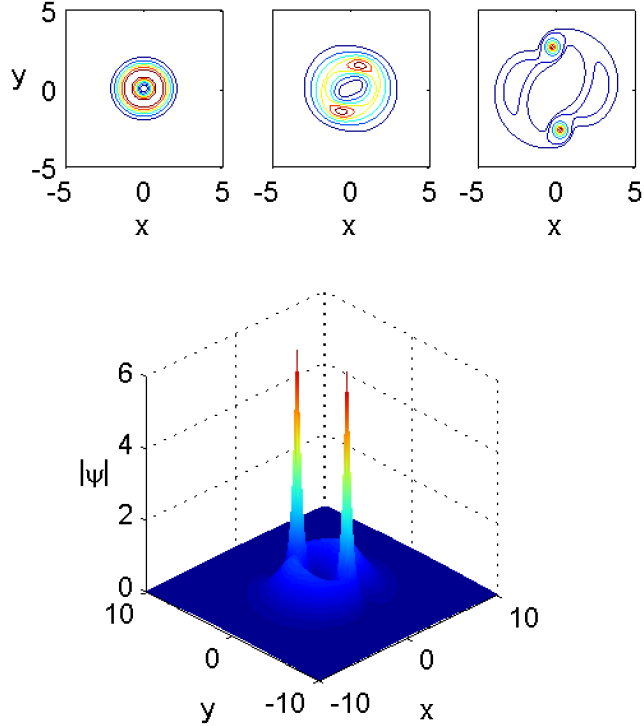}}
\caption{(color online) Solution of the NLS with input beam~(\ref{eq:G2_vortex_below_critical}).
Top: Levels set at~$z=0,0.5$ and~$z=1$ (from left to right) .
Bottom: Surface plot at $z=1$.
} \label{fig:G2_vortex_below_critical}
\end{center}
\end{figure}

Most studies on optical vortices considered stationary vortices.
Recently, there has been a growing interest in the dynamics of
collapsing vortices.  Berge et al. showed that for vortices with
input power~$P\approx I_c^{(m)}$, symmetry breaking noise causes the
vortex ring to breaks into~$2m+1$ filaments~\cite{Berge-05}. Vuong
et al. generalized this result for vortices with power larger
above~$I_c^{(m)}$~\cite{Luat-06}.  We now show that these azimuthal
instabilities can occur even  for vortices with dimensionless power
less than~$I_c^{(m)}$ and even less than the lower
bound~$p_{cr}(m)$. To do that, we solve the NLS with the slightly
elliptic Laguerre-Gaussian input profile
\begin{equation} \psi_0=\psi_0^{LG}\left(\sqrt{x^2+(1.05\cdot
y)^2}\right),\label{eq:G2_vortex_below_critical}
\end{equation}
with~$m=2$ and with input power equal to~$\frac34p_{cr}(m=2)$.
Although the power of this vortex beam is below~$p_{cr}(m)$, it
breaks into two filaments which subsequently undergo collapse, see
Figure~\ref{fig:G2_vortex_below_critical}.
This effect of symmetry-breaking is very different from the case of
peak-type non-vortex solutions, where deviations from radial
symmetry increase the critical power for
collapse~\cite{Elliptic-00}.   This is because peak-type solutions
collapse with the modulated Townes profile,
(i.e.,~$\psi\sim\frac1{L(z)}R\left(\frac{r}{L(z)}\right)$
where~$L\to0$ at the singularity) which is stable under azimuthal
perturbations, as was demonstrated experimentally and numerically
in~\cite{Moll-03}, and analytically in~\cite{Merle-03}. In contrast,
vortices collapse with a ring profile, which breaks into a ring of
filaments under azimuthal perturbations~\cite{Luat-06}. Since these
filaments do not collapse at the phase singularity point~$r=0$, each
filament can collapse with the Townes profile, hence with the
critical power~$p_{cr}=p_{cr}(m=0)<p_{cr}(m)$~\cite{footnote-1}.
Note, that these filaments continue to rotate around~$r=0$, so that
total helicity is preserved.

Our results are also relevant for beams which are not linearly
polarized. Let~$\psi_\pm$ be the amplitudes of the circular
components $\hat e_{\pm}=(\hat x\pm i\hat y)/\sqrt2$. The equation
for each circular component is
\[
i\frac{\partial\psi_{\pm}}{\partial z}+\Delta\psi_\pm+\frac23\left[|\psi_\pm|^2+2|\psi_\mp|^2\right]\psi_\pm=0.
\]
In the case of a pure circular polarization (CP)
state~($\psi_-\equiv0$), this equation reduces to
\begin{equation}
i\frac{\partial\psi_+}{\partial z}+\Delta\psi_++\frac23|\psi_+|^2\psi_+=0.
\label{eq:NLS_CP}
\end{equation}
Since the Kerr effect is smaller by a factor $2/3$ compared to the
NLS~(\ref{eq:NLS}) for a linear polarization state, the critical
power for collapse is larger by a factor
of~$3/2$~\cite{Explanation-1}. In particular, the lower bound for
the critical power of a CP vortex beam~$\psi_+=e^{im\theta}A_0(r)$
is given by
\[ p_{cr}^{CP}(m)=\frac32p_{cr}(m)\approx 6\sqrt3m.
\]
Similarly, consider the cases of radial polarization (RP)
\[
\psi^{RP}=A(r,t)[e^{i\theta}\hat e_- + e^{-i\theta}\hat e_+],
\] and azimuthal polarization (AP)
\[\psi^{AP}=iA(r,t)[e^{i\theta}\hat e_- - e^{-i\theta}\hat e_+].
\]
Since $|\psi_+|=|\psi_-|=|A|$, the equation for each component is
\[
i\frac{\partial\psi_{\pm}}{\partial z}+\Delta\psi_\pm+2 |\psi_{\pm}|^2 \psi_\pm=0.
\]
The Kerr effect is larger by a factor of $2$, hence the critical
power for collapse for each component is smaller by a factor
of~$\frac12$,
i.e.,~$p_{cr}(\psi_+)=p_{cr}(\psi_-)=\frac12p_{cr}(m=1)$. In
addition, the power of~$\psi^{AP}$ and~$\psi^{RP}$ is the sum of the
power of~$\psi_+$ and of~$\psi_-$.
Hence,\[p^{RP}_{cr}=p^{AP}_{cr}=p_{cr}(\psi_+)+p_{cr}(\psi_-)=p_{cr}(m=1)\approx
4.12 p_{cr},\] in agreement with recent numerical
simulations~\cite{AMI-07}.

In summary, we showed that the critical power for collapse of
radially-symmetric vortex beams is typically a few percent
above~$P_{cr}(m)= \frac{\lambda^2}{4\pi n_0n_2}p_{cr}(m)$
where~$p_{cr}(m)= \int_0^\infty R_{m}^2 \,rdr\approx4\sqrt3m$.
Deviations from radial-symmetry do not increase the critical power,
but rather lead to disintegration into collapsing non-vortex
filaments.

We thank Amiel Ishaaya for useful discussions.  This research was
partially supported by Grant No. 2006-262 from the United
States–-Israel Binational Science Foundation (BSF), Jerusalem,
Israel.  The research of Nir Gavish was also partially supported by
the Israel Ministry of Science Culture and Sports.
\bibliographystyle{apsrev}

\begin{thebibliography}{}
\bibitem{footnote-1} This dynamics is similar
to the case of non-vortex ring-type beams which, under azimuthal
perturbations, breakup into filaments, each of which collapse with
the Townes profile~\cite{SG_beams_collapse-06}.
\bibitem{Explanation-1}
This follows from the fact that if~$\psi$ is a solution of the
NLS~(\ref{eq:NLS}) than~$\psi_+=\sqrt{2/3}\psi$ is a solution
of~(\ref{eq:NLS_CP}).
\end{thebibliography}

\end{document}